\documentclass[final,3p,times]{elsarticle}

\usepackage{amssymb}
\usepackage{setspace}
\usepackage[T1]{fontenc}
\usepackage[utf8]{inputenc}
\usepackage[english]{babel}
\usepackage{csquotes}
\usepackage{ifthen}
\usepackage{url}
\usepackage{longtable}
\usepackage{placeins}
\usepackage{threeparttable}
\usepackage{tabularx}
\usepackage{makecell}
\usepackage{fourier} 
\usepackage{array}
\usepackage{multirow}
\usepackage{fancyhdr} 
\usepackage{graphics,graphicx,epsfig,ulem,float,caption,subcaption} 
\usepackage[labelfont=bf]{caption} 
\usepackage{amsmath,siunitx,cancel,dsfont,amssymb} 
\usepackage{lastpage} 
\usepackage{listings}
\usepackage[framed]{matlab-prettifier}

\usepackage{lipsum}

\usepackage{blindtext}
\usepackage{graphicx}

\let\oldhat\hat
\renewcommand{\hat}[1]{\oldhat{\mathbf{#1}}}

\renewcommand{\b}[1]{\textbf{#1}}

\newcommand{\fs}{\footnotesize}
\usepackage{lineno}
\usepackage{adjustbox}
\usepackage{pifont}
\usepackage{array}
\usepackage{enumitem}
\usepackage{ragged2e}

\journal{Computers in Biology and Medicine}

\begin{document}
\onehalfspacing
\begin{frontmatter}



\title{Practicality meets precision: Wearable vest with integrated multi-channel PCG sensors for effective coronary artery disease pre-screening}


\author[inst1,inst2]{Matthew Fynn}
\author[inst3]{Kayapanda Mandana}
\author[inst3]{Javed Rashid}
\author[inst1]{Sven Nordholm}
\author[inst1]{Yue Rong}
\author[inst2]{Goutam Saha}

\affiliation[inst1]{organization={School of Electrical Engineering, Computing and Mathematical Sciences (EECMS), Faculty of Science and Engineering},
                    addressline = {Curtin University},
                    city = {Bentley},
                    postcode = {6102},
                    state = {WA},
                    country = {Australia}}

\affiliation[inst2]{organization={Department of Electronics \& Electrical Communication Engineering},
                    addressline={Indian Institute of
Technology Kharagpur},
                    city={Kharagpur},
                    postcode={721302},
                    state={West Bengal},
                    country={India}}

\affiliation[inst3]{organization={Department of Cardiology},
                    addressline = {Fortis Healthcare},
                    city={Kolkata},
                    postcode={7007107},
                    state={West Bengal},
                    country={India}}

\begin{abstract}
The leading cause of mortality and morbidity worldwide is cardiovascular disease (CVD), with coronary artery disease (CAD) being the largest sub-category. Unfortunately, myocardial infarction or stroke can manifest as the first symptom of CAD, underscoring the crucial importance of early disease detection. Hence, there is a global need for a cost-effective, non-invasive, reliable, and easy-to-use system to pre-screen CAD. Previous studies have explored weak murmurs arising from CAD for classification using phonocardiogram (PCG) signals. However, these studies often involve tedious and inconvenient data collection methods, requiring precise subject preparation and environmental conditions. This study proposes using a novel data acquisition system (DAQS) designed for simplicity and convenience. The DAQS incorporates multi-channel PCG sensors into a wearable vest. The entire signal acquisition process can be completed in under two minutes, from fitting the vest to recording signals and removing it, requiring no specialist training. This exemplifies the potential for mass screening, which is impractical with current state-of-the-art protocols. Seven PCG signals are acquired, six from the chest and one from the subject's back, marking a novel approach. Our classification approach, which utilizes linear-frequency cepstral coefficients (LFCC) as features and employs a support vector machine (SVM) to distinguish between normal and CAD-affected heartbeats, outperformed alternative low-computational methods suitable for portable applications. Utilizing feature-level fusion, multiple channels are combined, and the optimal combination yields the highest subject-level accuracy and F1-score of 80.44\% and 81.00\%, respectively, representing a 7\% improvement over the best-performing single channel. The proposed system's performance metrics have been demonstrated to be clinically significant, making the DAQS suitable for practical use. Moreover, the system shows promise in post-procedural monitoring for subjects undergoing percutaneous transluminal coronary angioplasty (PTCA) or coronary artery bypass grafting (CABG), effectively identifying cases of restenosis following intervention.
\end{abstract}




\end{frontmatter}


\section{Introduction} \label{Intoduction}

    According to the World Health Organization, cardiovascular disease (CVD) stands as the foremost cause of morbidity and mortality globally, contributing to 31\% of all worldwide deaths in 2012 \cite{WHO2021}. Over the past few decades, developing countries have borne 80\% of CVD-related deaths, with atherosclerotic coronary artery disease (CAD) accounting for half of these reported cases \cite{cassar2009chronic}. This number is projected to reach 23.4 million by 2030 \cite{cassar2009chronic}. Coronary artery narrowing results from plaque deposition on the interior artery walls, leading to a diminished supply of oxygenated blood to the heart muscles. This process, known as stenosis, results in the gradual death of myocardial cells  \cite{smit2020pathophysiology}. Symptoms of CAD include chest pain, weakness, nausea, light-headedness, arm pain, and shortness of breath. A completely blocked coronary artery can cause myocardial infarction. It's noteworthy that CAD may also be present in patients with no symptoms \cite{kent1982prognosis}. Without proper treatment, CAD can progress to congestive heart failure and ischemic heart disease  \cite{desmondBook}. Therefore, early disease detection is crucial to prevent this irreversible state. Lifestyle changes or surgical measures, such as stent insertion and bypass surgery, can be performed to manage CAD  \cite{seung2008stents}. Presently, the gold standard for CAD diagnosis is coronary angiography  \cite{gibbons1999acc}, an invasive and expensive procedure that is often inaccessible to marginalized populations, especially in developing countries. Moreover, doctors typically recommend this technique only for patients experiencing symptoms, making early disease detection challenging. Alternative standard diagnostic tools include nuclear and exercise stress tests, which are costly and carry inherent risks \cite{akay2008dynamics, semmlow2007acoustic}. Unfortunately, a heart attack, stroke, or other extreme symptoms often become the first indicators of CAD long after its onset. Therefore, there is a pressing need for a low-cost and non-invasive method for pre-screening and early detection of CAD, which poses minimal risk to patients.
    \\
    
    Blood flow through a non-stenosed artery is streamlined and flows smoothly, exhibiting laminar characteristics \cite{thomas2017novel}. Plaque buildup causes disturbances leading to turbulent flow, characterized by high-frequency vibrations known as murmurs \cite{thomas2017novel}. The right coronary artery (RCA), left anterior descending artery (LAD) and left circumflex (Cx) are the primary arteries where blood flow becomes constricted due to stenosis \cite{gibbons1999acc}. Heart auscultation, a physician's interpretation of heart sounds, remains a common method for diagnosing cardiovascular disease (CVD) \cite{karnath2002auscultation}. Despite being cost-effective and non-invasive, heart auscultation is challenging to learn, requiring years of training. Many physicians are documented to have poor skills in this area; however, it remains a primary method for pre-screening in healthcare \cite{mahnke2004comparison,roy2002helping, roy2003paediatrician}. The literature reflects the growing popularity of computer-aided diagnosis, with machine-learning techniques adapted to identify normal and diseased patients through phonocardiogram (PCG) signals. Signal acquisition is often laborious and inconvenient, rendering mass screening impractical in terms of efficiency. CAD induces changes in the power spectral density (PSD) shape of PCG signals \cite{larsen2021spectral} due to murmurs. We aim to exploit these differences by extracting linear-frequency cepstral coefficients (LFCC) for classification, using signals acquired from a wearable vest integrated with PCG sensors. In this study, we challenge state-of-the-art DAQS and data collection methods, emphasizing ease and convenience for practical use. The main contributions of this paper are summarised below.
    
    \begin{itemize}
        \item The utilization of a wearable vest integrating six PCG signals from anterior positions and one PCG signal from a posterior position marks a novel approach in our study. To our knowledge, we are the first to implement this unique sensor combination and the first to investigate PCG signals acquired from the back, with performance demonstrating its benefit in CAD detection. 
        \item Our data collection method introduces unparalleled ease and convenience, surpassing current state-of-the-art techniques. This approach is directly applicable to practical use without specialist training, and preliminary results indicate clinical significance. 
        \item A low-computational approach of combining LFCC features from multiple PCG channels to classify CAD using a support vector machine (SVM) classifier renders the system suitable for portable applications.
        
        \item A novel system evaluation on subjects who have undergone percutaneous transluminal coronary angioplasty (PTCA) or coronary artery bypass grafting (CABG) indicates the potential for post-procedural monitoring. 
    \end{itemize}

     The subsequent sections of this paper are structured as follows: Section \ref{Background} provides a concise background on the acquired biosignals, Section \ref{Literature} critically evaluates the literature on CAD detection, Section \ref{DAQS} details the DAQS and novel data collection methodology, Section \ref{DPM} covers pre-processing and feature extraction for classification, and the results are discussed in Section \ref{results}. Comparisons are made to other feature types in Section \ref{compare}, and in Section \ref{Additional}, additional experimentation on PTCA and CABG subjects is conducted. Comparisons to existing studies are presented in Section \ref{compare to lit}, and a discussion of practical implementation is delivered in Section \ref{practical}. The conclusion and future direction are in Sections \ref{Summary} and \ref{Future}, respectively.

    \section{Background} \label{Background}

    \subsection{Phonocardiogram}
\begin{figure}[ht]
	    \centering
	    \includegraphics[width = \textwidth]{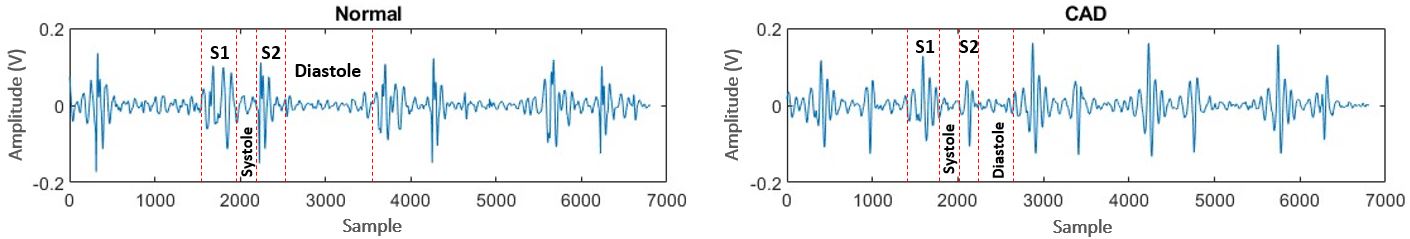}
	    \caption{PCG of a normal and a typical CAD subject with two blocked vessels}
	    \label{cad nor}
	\end{figure}

    As venous blood returns to the heart's right atrium from the body's organs, it is sent to the lungs for reoxygenation from the right ventricle and then circulates back through the pulmonary veins to the left atrium. The oxygenated blood is subsequently pumped from the left ventricle through the aortic valve to be distributed among the body's organs. Additionally, blood is supplied to the heart through coronary arteries branching from the aorta. Heart sounds are generated throughout this process, with PCG signals providing acoustic representations. Vibrations resulting from the closure of the atrioventricular valves produce the S1 sound, marking the onset of systole when the heart contracts to propel blood through arteries from its chambers. The S2 sound, on the other hand, originates from the closure of the aortic and pulmonic valves at the beginning of diastole, when the cardiac muscle relaxes, facilitating the filling of the heart chambers with blood. \cite{cardiovascular}. The prevalence of CAD can cause variations in the PCG signal, previously stated as murmurs \cite{thomas2017novel}. These are predominately noticed during systole and diastole as the blood flows in and out of the heart chamber, indicating the turbulent flow of blood \cite{cardiovascular}. Figure \ref{cad nor} displays a  PCG from a normal and CAD-affected subject with two blocked arteries, with an annotated S1, systole, S2 and diastole. Murmurs present in heartbeats are not always due to CAD. Other valvular diseases can cause these high-frequency sounds. Hence, it is important to distinguish these other sources when detecting CAD to prevent misdiagnosis. Murmurs associated with valvular pathology exhibit greater intensity than those arising from CAD. Consequently, the primary emphasis of this study is directed towards detecting murmurs specifically attributable to CAD, acknowledging the heightened complexity inherent in this investigative task. Subjects with identifiable murmurs from pre-screening should be referred to other standard diagnostic tools mentioned previously. 

    \section{Literature Survey}\label{Literature}
    
    Since 1992, numerous studies have been published in the literature relevant to computer-aided heart disease detection.  Accuracy, sensitivity, and specificity stand as the three primary quantifiable metrics indicating the efficacy of a machine-learning classifier. Factors such as dataset size, quality, feature selection, and the choice and number of features significantly influence the design of an optimal classifier \cite{wolpert1997no}. In some cases, a mixture of multiple classifier types can be fused to produce better results \cite{duda2006pattern}. This literature survey aims to delineate the evolution of literature concerning CAD detection, commencing with single-channel PCG and progressing to multi-channel PCG analysis. Subsequently, a critical evaluation of the DAQS and data collection methods is provided, emphasizing the novelty of our proposed method.

    \subsection{Single-Channel PCG}

   Makaryus \textit{et al.} \cite{makaryus2013utility} used an advanced digital electronic stethoscope to capture heartbeats measured at nine precise positions on the chest for 40 seconds each. Patients were seated, and an elastic strap held the stethoscope in place during recording, with a separate acoustic sensor monitoring background noise and electrocardiographic leads attached for heartbeat segmentation. A generated microbruit score from 142 non-CAD and 19 CAD subjects was used to train a logistic regression model. The accuracy, sensitivity, and specificity achieved were 61.5\%, 89.5\%, and 57.7\%, respectively. Schmidt \textit{et al.} \cite{schmidt2015acoustic} investigated frequency and non-linear features from 133 patients comprising 63 CAD and 70 normal. 8-second bedside recordings were made with an electronic stethoscope placed on the fourth intercostal (IC) space left of the sternal border from normal-breathing patients. The accuracy, sensitivity, and specificity achieved were 68.4\%, 72\%, and 65.2\%, respectively, using features from five overlapping frequency bands. Li \textit{et al.} \cite{li2020fusion} investigated the fusion of handcrafted and deep learning features on a single-channel PCG. 120 CAD and 55 non-CAD patients (confirmed via angiography) were instructed to lie in a temperature-controlled room (25$^{\circ}$) for 15 minutes between 2-6 pm before 5-minute heartbeat signals were recorded by a piezoelectric sensor placed in the third IC space, left of the sternal border. An accuracy, sensitivity, and specificity of 85.03\%, 90.50\% and 73.09\%, respectively, were reported using handcrafted features. The accuracy increased to 90.4\% when fused with deep learning features. The authors of this study did not specify how breathing sounds affected the PCG recordings. In future studies, it would be of interest to observe how the classifier performs when the subjects are in a breath-held state, as it is clear that breathing noise corrupts PCG signals \cite{fynn2022coherence}. Huang \textit{et al.} \cite{huang2022customized} collected data from 206 CAD and 348 non-CAD subjects. A high-cost, non-invasive cardiovascular detector (DR-A-1) acquired signals from the sternum parallel to the third rib space from supine subjects placed in a temperature-controlled room 5 minutes prior with low noise. A two-branch CNN and LSTM model resulted in 96.05\% accuracy, where high-cost equipment and precise sensor location contributed to the high performance. Iqtidar \textit{et al.} \cite{iqtidar2021phonocardiogram} designed a classifier to identify the severity of CAD in patients. The dataset consisted of 19 single vessel CAD (SVCAD), 24 double vessel CAD (DVCAD), 35 triple vessel CAD (TVCAD), and 75 normal subjects. An expert cardiologist labeled the PCG signals acquired from the mitral position in a clinical environment. This is a disadvantage since the gold standard angiogram did not confirm the subject diagnosis. Mel-Frequency Cepstral Coefficients (MFCC) and 1D-Adaptive Local Ternary Patterns produced binary and multiclass accuracies of 98.3\% and 97.2\%, respectively. Although the results were high, only the CAD patients were collected in a hospital environment, and multiple samples per subject were taken. As subject-level accuracy was not reported, it appears that samples from the same subject could appear in both the training and test sets, contributing to data leakage and high performance. Many studies have analyzed the PhysioNet/CinC Challenge 2016 database \cite{nia2024abnormal,rong2023wearable,marocchi2023abnormal, maity2023transfer}. However, subjects with valvular pathologies, as well as CAD, are also included. Here, data was collected in both clinical and non-clinical environments.
    
    \subsection{Multi-channel PCG}
    Pathak \textit{et al.} \cite{pathak2020detection} collected multi-channel PCG signals from 40 CAD and 40 normal subjects. Four stethoscopes with condenser microphones placed in the plastic tubing were taped to the left second IC space, the left fifth IC space on the mid-clavicular line, the left fourth IC space and the left fourth IC space on the midaxillary line. Heart sounds were recorded for 10 seconds in the supine position while the patient was breath-held. Each recording was split into three epochs containing two full heart cycles; thus, epoch-level and subject-level metrics were reported. The synchrosqueezing (SST) transform of each epoch was obtained, which is computationally expensive. Sub-band entropy features were extracted from the SST matrix across different time frames, and 5-fold cross-validation with SVM-linear was utilised for classification. A subject-level accuracy of 84.81\% was reported.  Liu \textit{et al.} \cite{liu2021detection} collected synchronous PCG signals from 21 CAD and 15 non-CAD subjects using five electronic stethoscopes placed in precise positions for 5 minutes. After preprocessing, the signals were segmented to produce 533 CAD and 438 non-CAD samples. Entropy features from single channels and cross-entropy features from paired channels resulted in 90.92\% accuracy; however, all CAD subjects had left anterior descending stenosis, favouring classification as similar characteristics will be present in the data. In real-life scenarios, CAD patients with different affected vessels will be present. For example, the system developed in this study may not generalize to CAD patients with stenosis in their right coronary artery.

    \subsection{Critical Evaluation of Literature: Common Practical Limitations}
    

    The reviewed studies and others available in the literature share a common objective: to assist physicians in the early detection of CAD. However, implementing these studies in practical scenarios poses challenges. The data collection process tends to be tedious and inconvenient, making mass screening impossible in an efficient manner. In each study, sensors are precisely positioned on subjects, which requires specialist training, typically using tape \cite{pathak2020detection, pathak2022ensembled}, elastic bands \cite{makaryus2013utility}, or without specified attachment methods \cite{schmidt2015acoustic, li2020fusion, iqtidar2021phonocardiogram, huang2022customized, liu2021detection, li2019dual, dong2023non}. It's reasonable to assume that studies involving multi-channel PCG data without explicit sensor attachment details had the sensors taped to the subject's chest, extending the time needed to prepare each patient. Single-channel PCG studies likely employed hand-held sensors, potentially introducing friction noise due to tremors. Subjects are typically required to be in a supine position, \cite{schmidt2015acoustic, li2020fusion, huang2022customized, pathak2020detection, pathak2022ensembled, li2021integrating, dong2023non}; thus, a hospital bed is required. In some studies, subjects are instructed to remain in a temperature-controlled and quiet room for 5-10 minutes during a specified time of day \cite{li2020fusion,huang2022customized, li2021integrating, dong2023non}, which may not be feasible in rural or busy settings. However, some studies have collected data in a clinical setting \cite{iqtidar2021phonocardiogram}, introducing realistic noise to the collected signals. It's common to observe the use of high-cost equipment for data acquisition. For example, \cite{li2020fusion,li2021integrating,dong2023non} employs a CVFD-II cardiovascular function detector, while \cite{huang2022customized} utilizes a DR-A-1 detector. This enables the acquisition of high-quality and clean signals. Recent studies have delved into multi-class classification, whether for determining the number of blocked vessels \cite{iqtidar2021phonocardiogram} or the severity of CAD \cite{dong2023non}. However, since this study aims to develop a pre-screening device for asymptomatic individuals, we have chosen to detect all stages of CAD using a two-class system, even though our current CAD participants are symptomatic. This decision stems from the understanding that anyone flagged by the system will be referred to gold standard methods, where a more comprehensive diagnosis can be conducted.\\

    While the outlined data collection methods are indeed tedious and inconvenient, they contribute significantly to achieving high-performance metrics. A precise set-up of environment, subject positioning, sensor location, and high-cost equipment reduces the variation in signals obtained from different subjects, allowing extracted features to be relatively consistent. This ideal scenario, where signal variation between subjects is solely attributed to biological factors while all other variables are tightly controlled, greatly facilitates the classification process by reducing the presence of noise within the system. However, this approach may not be practical in real-world environments such as hospitals, rural clinics, or home settings. Descriptions of sensor placement can be subjective among practitioners, and home users may struggle to understand the terminology used to describe sensor positioning. Patient throughput will also be reduced if tape is needed to ensure precise positioning. Environmental conditions, particularly temperature, may also vary based on geographical location or setting. For instance, a home user may experience warmer conditions during the summer than a patient in a hospital ward, making temperature control unrealistic. Furthermore, finding a quiet location may not always be feasible, and low-socioeconomic areas may lack access to high-cost DAQS. The objective of this study is to attain acceptable accuracy suitable for practical and portable applications, employing a convenient, low-cost, and non-invasive data collection process accessible to both physicians and the common layperson. There will be no requirement for tedious patient preparation or adherence to strict environmental conditions. The criteria for acceptable accuracy in practical use are defined in \cite{power2013principles}, where the average between sensitivity and specificity must exceed 75\%. The next section of this paper describes the newly designed DAQS and the methodology of multi-channel PCG acquisition.

    \section{Data Aquisition System} \label{DAQS}
    
    \subsection{Hardware} \label{hardware}
\begin{figure}[ht]
    \begin{subfigure}[t]{0.5\textwidth}
        \includegraphics[width=6cm]{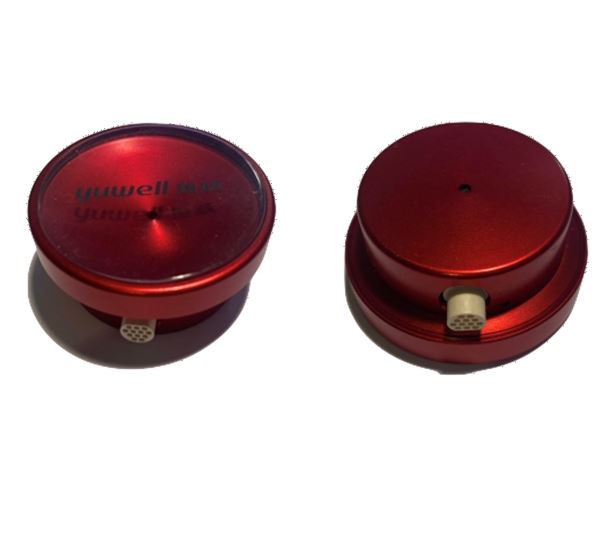}
		\caption[]{}
		\label{steth}
    \end{subfigure}
    ~
    \begin{subfigure}[t]{0.5\textwidth}
        \centering
        \includegraphics[width = 8cm]{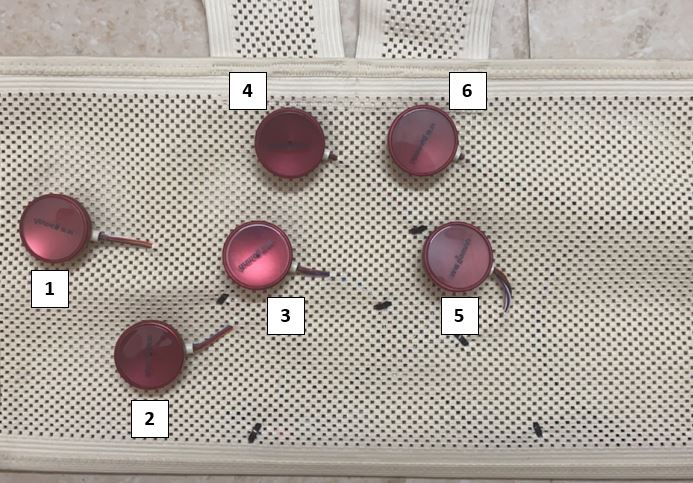}
		\caption[]{}
		\label{openvest}
    \end{subfigure}
    \caption{\centering \textbf{(a)} Stethoscope \textbf{(b)} Wearable vest fitted with electronic stethoscopes}
    \end{figure}

    To bridge the research gap and achieve clinically significant CAD classification performance with practical ease and convenience, we are utilizing a DAQS that integrates up to seven electronic stethoscopes into a vest worn by the subject to capture heart sounds. Multiple PCG channels can offer a richer representation of heart abnormalities and provide more information than a single auscultation site \cite{nogueira2022can}. Among the stethoscopes, six are positioned on the front of the body, while one is placed on the back. To our knowledge, no previous studies have analyzed this multi-channel configuration, especially on the posterior side. The stethoscopes are fixed to the vest and are non-adjustable. At the front of each stethoscope, located beneath the diaphragm, is the heart microphone (HM). Its purpose is to capture the acoustic signature of the heart. As the diaphragm makes contact with the skin, any movements result in pressure variations within the stethoscope-diaphragm cavity, detected by the HM as heart sounds that penetrate towards the skin surface. All signals captured from the stethoscopes are channelled to a central data collection board (hub) via custom-made cables for further signal conditioning before synchronous digitalization. This setup employs a 24-bit sigma-delta converter with a sampling frequency of 7.812 kHz. The hub is powered through a USB connection, and a graphical user interface (GUI) was developed using MATLAB 2023a. The final recordings are saved and processed on a laptop equipped with an Intel(R) Core(TM) i7-10510U processor, 16 GB of RAM, and running Windows 10 with a processor speed of 2.3 GHz. The multi-signal data is stored as a multitrack .wav file and includes all stethoscope HM recordings. Figure \ref{steth} displays the upward and downward-facing electronic stethoscope. This is the wearable vest's fundamental building block, as presented in Figure \ref{openvest}. The expected cost of the vest will be targeted at consumer-level markets, aiming to be comparable to other at-home health monitoring devices. The hardware was obtained from a private organization. The sensor placement and data collection protocol, described in the next sub-section, are novel aspects of this study.

    \subsection{Data Collection Methodology}
    
    Between May and June 2023, multi-channel PCG data were gathered at Fortis Hospital, Kolkata, using the wearable vest. In this preliminary investigation, 80 male subjects participated, comprising 40 CAD patients, 32 non-CAD patients, and 8 male nurses aged 21-29, assumed to be in good health. We can safely assume the control subjects do not have CAD, as the risk of developing CAD is significantly higher for males over 45 years of age, and these subjects have no risk factors \cite{hajar2017risk}. Subjects with valvular pathologies or those who had undergone PTCA or CABG procedures were excluded from this initial phase of the study. Verification of CAD and non-CAD patients was conducted through gold standard coronary angiography, wherein a subject is classified as CAD if either the RCA, LAD, or Cx exhibits more than 50\% stenosis according to standard guidelines \cite{gibbons1999acc}. Thus, an evenly balanced dataset was obtained by including the 8 male nurses in the non-CAD group. 
    Within the CAD group, 14 subjects present with single-vessel CAD (SVCAD), 14 with double-vessel CAD (DVCAD), and 12 with triple-vessel CAD (TVCAD), each exhibiting varying degrees of stenosis. Given the nature of this pre-screening device, our objective is to detect CAD at all stages and in all arteries. The average (standard deviation) age of the CAD group is 59.73 (8.02) years, while that of the Normal group is 49.7 (18.8) years. The average body mass index (BMI) for the CAD and Normal group is 24.62 (4.19) kg/m$^2$ and 23.92 (3.03) kg/m$^2$, respectively.\\
    
    Since the stethoscopes are affixed to the vest, there is inherent variation in their placement among subjects due to differences in body types and the fitting process conducted by doctors and nurses. As a result, signal characteristics from a particular stethoscope may exhibit slight variations across subjects that are unrelated to biological variables. This variability arises from the simplification of the data acquisition process, which is the primary focus of this study. However, efforts were made to mitigate this issue by providing multiple vest sizes: medium (M), large (L), and extra-large (XL), with only L and XL vests utilized in this study. Despite the presence of placement variation, the seven stethoscopes are positioned to capture information from approximately the same areas of the chest and back. Specifically, four stethoscopes are placed on the left-hand side, two on the right-hand side, and one on the back. The placement of the vest was carefully selected under the assumption that subjects would be seated upright or standing during data acquisition, thus eliminating the necessity for a hospital bed. Channel 1 is positioned approximately over the midaxillary line on the left fourth intercostal (IC) space; channel 2 is situated around the midclavicular line below the fifth IC space near the apex; channel 3 is positioned near the left fourth IC space; channel 4 is located near the left second IC space; channel 5 is placed on the opposite side of the sternum from channel 3; channel 6 mirrors the placement of channel 4 on the right-hand side; and channel 7 is positioned on the left side of the back along the same horizontal line as channel 4. \\

\begin{figure}[ht]
	    \centering
	    \includegraphics[width = \textwidth]{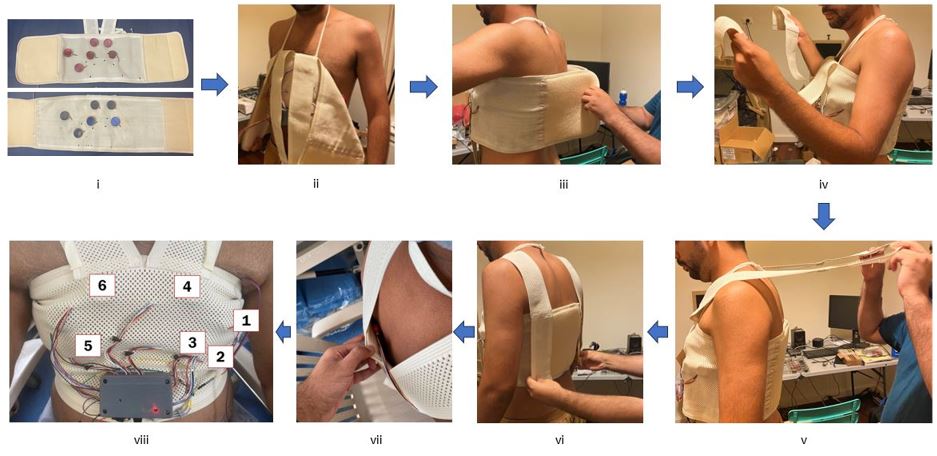}
	    \caption{Steps taken to fit the vest on a subject. i) Select the appropriate vest size based on body dimensions. ii) Place the neck strap to ensure the middle of the vest aligns with the sternum. iii) Wrap the vest tightly around the body, securing it firmly with stitched-on Velcro. iv, v, vi) Wrap the shoulder straps over the shoulders and attach them to the back of the vest using Velcro. vii) Position the back stethoscope at the rear of the vest. viii) Ensure that all stethoscopes make clean contact with the skin. }
	    \label{vest fitting}
	\end{figure}

    Fitting the vest onto a shirtless subject can be done individually or with the assistance of another person and requires no specialist training. Once the appropriate-sized vest is selected, a strap is placed around the neck to prevent the vest from slipping and to ensure alignment with the sternum. The vest is then wrapped around the body and securely fastened with Velcro on the back. Two vertical straps are then wrapped over the shoulders and affixed to the back of the vest using Velcro strips. This further tightens the vest to ensure optimal contact between the stethoscopes and the skin. Finally, the back stethoscope is positioned at the rear of the vest. If necessary, the stethoscopes are adjusted through the vest to ensure proper contact with the skin. The entire vest fitting process can be completed within 30 seconds. Figure \ref{vest fitting} illustrates the fitting process of the vest, culminating in the vest being worn by a subject, with stethoscope numbering corresponding to Figure \ref{openvest} and the channel placement description. Additionally, the figure depicts the hub enclosed in a plastic casing for protection. Ideally, the stethoscopes would capture data between adjacent ribs to minimize extra dampening. However, due to subject-to-subject variability and the fixed positioning of stethoscopes on the vest, achieving this ideal scenario for all cases is not feasible. This introduces further signal variation between subjects, representing an additional consequence of the user-friendly system design. \\

\begin{figure}[ht]
        \centering
        \includegraphics[width=\textwidth]{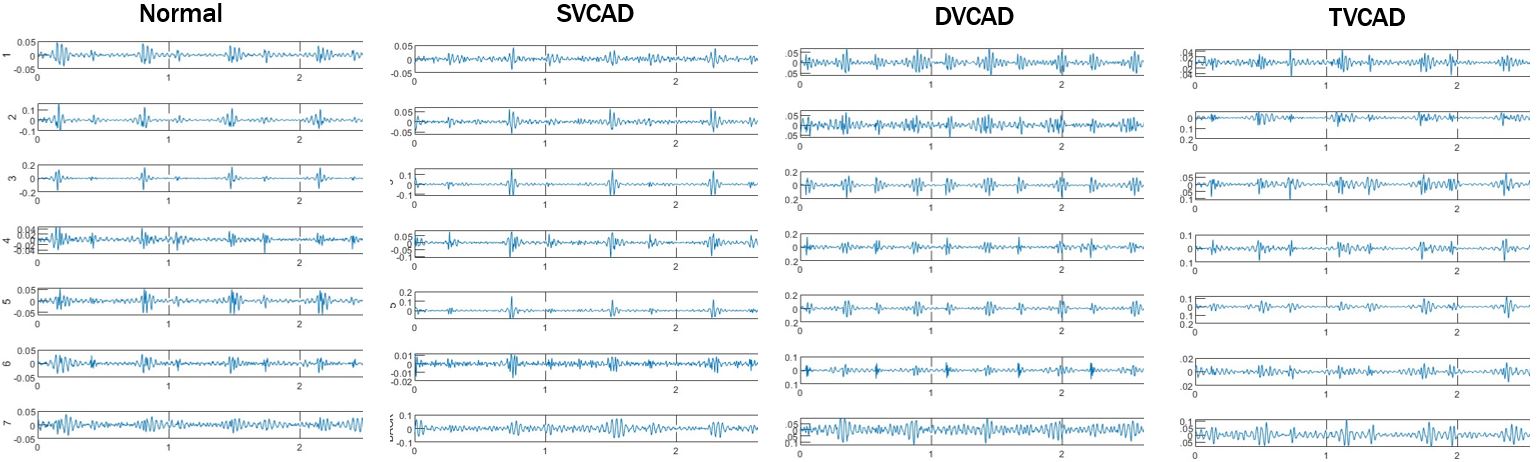}
        \caption{Typical signal of non-CAD, SVCAD, DVCAD, and TVCAD subject high-pass filtered with 20 Hz cutoff frequency}
        \label{recording1}
    \end{figure}

     All recordings took place in the angiography ward of Fortis Hospital, Kolkata, where patients undergo preparation for coronary angiography. The ward can accommodate up to 12 patients and their family members simultaneously, with 6-10 nurses and 2-3 ward staff working in the same area. This clinical setting introduces various types of noise into the system. Observed noise sources include, but are not limited to, talking, doors closing, privacy curtains opening and closing, electric shavers, footsteps, ringing phones, and the sound of tap water flowing into a metal basin. Occasionally, construction was taking place in neighbouring rooms or buildings. These types of noise sources will be expected in real-life scenarios. Patients were invited to participate in the study before undergoing angiography. Informed consent was obtained from all patients and healthy volunteers, and data collection adhered to the code of ethics for conducting research on human subjects as outlined in the Helsinki Declaration. \cite{wma}. Following the fitting of the vest on the subject, a brief 5-10 second test recording was performed to verify that the stethoscopes captured readable signals. This test ensured that no loose connections would compromise the final recording quality. Subjects were instructed to hold their breath for the duration of the 10-second recording. Once measurements were completed, the vest was removed in reverse order as depicted in Figure \ref{vest fitting}. Figure \ref{recording1} compares breath-held recordings obtained from a Normal, SVCAD, DVCAD, and TVCAD subject. In a practical scenario, assuming a 10-second recording is obtained, the entire process of fitting the vest, testing the connections, recording, and removing the vest can typically be completed within 1-2 minutes. This emphasizes the convenience of the system, as no patient preparation is required before fitting the vest. To the best of our knowledge, no data collection methodology described in the literature offers this level of ease and convenience. Table \ref{lit compare} provides a comparison of this study's DAQS with others documented in the literature.

    Other wearable designs for PCG signal acquisition were also explored. For example, chest straps were considered since they are similar to heart rate monitors and can be fitted with multiple stethoscopes. However, multiple straps would be needed to ensure sufficient coverage, increasing the time required for signal acquisition and causing discomfort to subjects. Ultimately, we chose the wearable vest due to its good chest coverage and ease of use. It is comfortable for subjects, and the snug fit reduces motion artifacts. Moreover, the design is scalable and practical. It can be adjusted to accommodate different body shapes and sizes, making it suitable for use in diverse patient populations. The simplicity of the design also aids in mass production and deployment.

        \begin{table}[!h]
        \centering
        \caption{\textbf{DAQS Comparison with existing studies}}
        \begin{adjustbox}{width=1\textwidth}
        \label{lit compare}
        \begin{tabular}{c|ccccccc}
            \hline \hline
            Study & \makecell{Acquired \\ biosignal} & \makecell{Patient \\ preparation} & \makecell{Sensor 
            \\ positioning} & \makecell{Sensor \\ attachment} &Environment & \makecell{Subject \\ position} & Cost  \\
            \hline
            \makecell{Makaryus \textit{et al.} \\ \cite{makaryus2013utility} (2013)} & \makecell{Single-\\Channel \\ PCG} & \makecell{4 electro-\\cardiographic\\ leads attached for\\ heartbeat \\segmentation} & \makecell{Nine precise \\ positions (multiple \\single-channel \\ recordings)} & Elastic Strap & \makecell{Environmental \\ Noise Present} & Seated & Low*\\
            &\\
            \makecell{Schmidt \textit{et al.} \\\cite{schmidt2015acoustic}  (2015)} & \makecell{Single-\\Channel \\ PCG} & \makecell{Not Specified}& \makecell{Precise - \\ fourth left \\ intercostal space \\ left of sternal \\ border} & Hand-held & Clinical & Supine & Low \\
            &\\
            \makecell{Li \textit{et al.} \\ \cite{li2020fusion} (2020)} & \makecell{Single-\\Channel \\ PCG} & \makecell{Lie in temperature- \\controlled room \\(25 $\pm$ 3$^\circ$C) \\for 15 min between \\2 and 6 p.m. \\in supine position}&\makecell{Precise - \\ third left \\ intercostal space \\ left of sternal \\ border} & Hand-held* & \makecell{Quiet and \\ temperature- \\ controlled \\room} & Supine & High* \\
            &\\
            \makecell{Iqtidar \textit{et al.} \\ \cite{iqtidar2021phonocardiogram} (2021)}& \makecell{Single-\\Channel \\ PCG} & Not Specified & \makecell{Precise - \\ mitral position} & Hand-held* & Clinical & \makecell{Not \\Specified} & Low\\
            &\\
            \makecell{Huang \textit{et al.} \\ \cite{huang2022customized} (2022)} & \makecell{Single-\\Channel \\ PCG} & \makecell{Lie in temperature- \\controlled room \\(25 $\pm$ 3$^\circ$C) \\for 15 min in \\ supine position}& \makecell{Precise - \\ third rib space \\ on left side} & Tape*&\makecell{Quiet and \\ temperature- \\ controlled \\room}& Supine&High \\
            &\\
            \makecell{Pathak \textit{et al.} \\ \cite{pathak2020detection}  (2020)}& \makecell{Multi-\\Channel \\ PCG} & \makecell{Not Specified} & \makecell{Four precise \\ positions} & \makecell{3M Transpore \\ Tape} & Quiet Room & Supine & Low*\\
            &\\
            \makecell{Liu \textit{et al.} \\ \cite{liu2021detection} (2021) }& \makecell{Multi-\\Channel \\ PCG} & Not Specified & \makecell{Five precise \\ positions} & \makecell{Tape*} & Quiet Room* & \makecell{Not\\ Specified} & High*\\

            &\\
            This Study & \makecell{Multi-\\Channel \\ PCG}& \makecell{No preparation \\required}& \makecell{No precise\\positing due to \\fixed stethoscope-\\vest placement}&Wearable Vest & Clinical& \makecell{Seated or \\standing}&Low\\

            \hline \hline
            \multicolumn{8}{l}{*Indicates that the relevant detail was not specified in the paper and an educated assumption has been made} \\
        \end{tabular}
        \end{adjustbox}
        
   \end{table}
    \newpage
    \section{Data Processing Methodology} \label{DPM}

    \subsection{Classification Model and performance metrics}

    SVM has demonstrated strong performance in the classification of PCG biosignals \cite{pathak2020detection}, hence it has been adopted in this study. The objective is to achieve an average sensitivity-specificity (Sens-Spec) score of 75\% or higher, as specified by \cite{power2013principles}, to establish the suitability of the system for practical use. Additionally, accuracy (Acc) and F1-score, which provide comprehensive performance metrics, are reported. A stratified 5-fold cross-validation scheme is employed over 20 iterations for classification. Subject indices are randomly shuffled for each iteration before being divided into five folds. Thus, the results presented in this section represent the mean obtained over 100 models. SVM with a Radial Basis Function (RBF) kernel is implemented using the LIBSVM library in MATLAB \cite{chang2011libsvm}. To help justify the use of SVM in this study, relevant comparisons have been conducted with other classifiers, including k-Nearest Neighbors (k-NN), Artificial Neural Network (ANN), and Random Forest (RF). 

    \subsection{Pre-Processing} \label{preprocessing section}
    
\begin{figure}[ht]
	    \centering
	    \includegraphics[width = \textwidth]{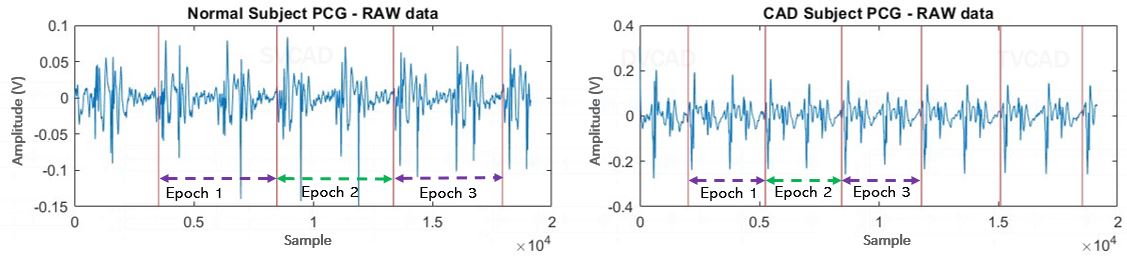}
	    \caption{Each 10-second signal is segmented into three epochs containing two full heart cycles. Only the first three epochs are used in this study.}
	    \label{hand seg}
	\end{figure}

    Each 10-second recording obtained from the 40 CAD and 40 non-CAD subjects underwent low-pass filtering at 1000 Hz using an 8th-order Butterworth filter before resampling at 2 kHz. The signal was then segmented into epochs, each containing two complete heart cycles to increase the number of data points. An epoch began slightly before the first heart sound (S1) of the first cycle and ended at the end of diastole of the second heart cycle. For this study, the first three epochs of each subject were utilized, starting from the second heartbeat. Consequently, 840 CAD and 840 normal PCG epochs (40 subjects per class $\times$ 3 epochs $\times$ 7 channels) were available. Manual segmentation was performed on channel 3 as its position was directly above the heart on the left side of the chest. The same indices were applied to segment all other channels to maintain the natural time delay of events captured from each transducer. Figure \ref{hand seg} exhibits the segmented PCG signal from channel 3 alongside Normal and CAD patient. During cross-validation, each subject's three epochs consistently appeared in the same fold to prevent data leakage. Majority voting (2 out of 3) was employed to compute subject-based metrics from epoch-based metrics. For instance, a subject with two out of three epochs classified as CAD (Normal) would be categorized as CAD (Normal). Epoch-based metrics provide insight into the model's performance on smaller time segments and its ability to detect specific patterns. A more comprehensive evaluation on a larger time scale is provided by subject-based metrics, which are critical for real-world applications that require a single decision per patient. This approach also enhances robustness to noise and short-term anomalies. If one epoch is corrupted, such as by a ringing phone or slamming door, and produces a false prediction, the subject can still be correctly classified based on the other epochs. Additionally, data bias concerns are addressed. If subject-based metrics underperform compared to epoch-based metrics, it may suggest that the model is overfitting to specific fragments of the data and not generalizing over the entire signal.\\

    Following segmentation, each epoch was z-normalized as per Equation \ref{z-norm}:

    \begin{equation}
        x_\mathrm{{z-norm}}(n) = \frac{x_\mathrm{{resampled}}(n) - \mu_x}{\sigma_x}, \; n = 1,2,...,N 
        \label{z-norm}
    \end{equation}

    \noindent where $x_{\mathrm{resampled}}$ is the segmented epoch after filtering and resampling, $\mu_x$ ($\sigma_x)$ is the average (std) epoch value, $N$ is the total number of samples in the epoch, and $x_\mathrm{{z-norm}}$ is the z-normalized signal. Z-normalization is essential due to variations in signal amplitude caused by different body characteristics, such as chest thickness. Following segmentation, the average (standard deviation) epoch length was found to be 1.79 (0.311) seconds, corresponding to 3583 (623) samples. The minimum epoch length observed was 1.05 seconds (2111 samples), while the maximum epoch length recorded was 2.65 seconds (5299 samples).
    
    \subsection{Power Spectral Density (PSD) Feature Formulation}
\begin{figure}[ht]
        \includegraphics[width=\textwidth]{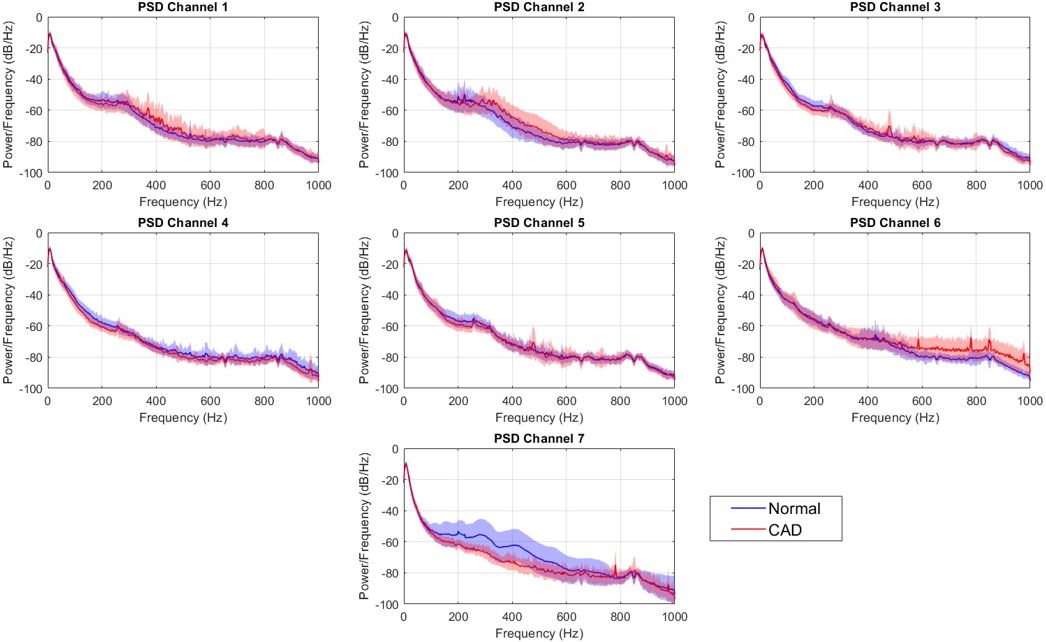}
        \caption{Average PSD of 120 CAD and 120 normal epochs from each channel. The shaded regions represent the standard deviation}
        \label{psd}
    \end{figure}

    For each channel, the PSD was computed for each epoch using the P-Welch method in MATLAB. A Hanning window, represented by Equation \ref{hanning}, consisting of 1024 samples (approximately 0.5 seconds) with 50\% overlap, was employed to reduce spectral leakage:
    \begin{equation} \label{hanning}
        w(n) = 0.5\left(1-\cos{\left(\frac{2\pi n}{N}\right)}\right) ,\; \; \; 0 \leq n \leq N
    \end{equation}
    
    \noindent where $n$ is the relevant sample and $N$ is the total number of samples in the frame ($N=1024$). Figure \ref{psd} displays the average and standard deviation of 120 CAD and 120 normal epoch PSDs from each channel. Discrepancies are noticeable in each channel across certain frequency bands. For instance, in channel 1, the normal curve is elevated between 100-250 Hz, whereas in channel 2, the CAD curve is higher between 300-600 Hz. Notably, the stethoscope on the subject's back exhibits significant disparities, with the normal curve being greater between 100-760 Hz. However, the spread between normal subjects in certain regions was nearly three times as large as that observed among CAD subjects. All PSD curves were segmented into sub-bands to assess the feasibility of these differences in classification. Sub-bandwidths (SBW) ranging from [5.86, 11.72, 17.58, 23.44, 29.30, 35.16, 46.88, 58.6] Hz were examined, alongside total bandwidths (TBW) of 0 - [300, 400, ...1000] Hz. These non-integer SBW selections ensure that each sub-band contains a consistent number of frequency bins. The MATLAB function \textit{trapz} was employed to calculate the average power within each band, which served as features in the classification model. Following feature extraction, a Minimum-Redundancy Maximum-Relevance (MRMR) filter is applied to rank the features \cite{peng2005feature}. This ranking ensures that the selected features exhibit minimal redundancy with respect to other features while maximizing their correlation with the class. The FEAST toolbox \cite{brown2012conditional} was employed to implement this process. Subsequently, various feature dimensions comprising the highest-ranked features were evaluated using an incremental search approach, with a step size of 2. 
    
    Single-channel analysis was initially conducted to determine the optimal SBW, TBW, and feature dimension (FD), resulting in the highest performance. The same process was conducted for multi-channel analysis. In multi-channel analysis, features from respective single channels were concatenated before passing through the MRMR filter and subsequent SVM model. For instance, when testing an SBW of 5.86 Hz over a TBW of 0-1000 Hz on channel combination 1-3-6, features were extracted from each single channel and merged into a single feature vector. The channel combinations that demonstrated the best performance are presented.

   \subsection{Linear-Frequency Cepstral Coefficient Features (LFCC)}
    Inherent to the PSD function is the loss of all temporal information, as power across various frequency bands is averaged over time. Thus, localised events occurring within an epoch are understated. To address this limitation, an investigation into LFCC feature extraction was conducted. LFCC aims to extend the PSD formulation by analyzing a compact spectral envelope representation and introducing temporal changes into the classifier by dividing the epoch into frames. Furthermore, the motivation behind this approach stemmed from the opportunity to represent better the human ear's logarithmic sensitivity to sound intensity, drawing parallels to heart auscultation with an acoustic stethoscope.

    \subsubsection{LFCC Feature Formulation}
\begin{figure}[ht]
	    \centering
	    \includegraphics[width = 13cm]{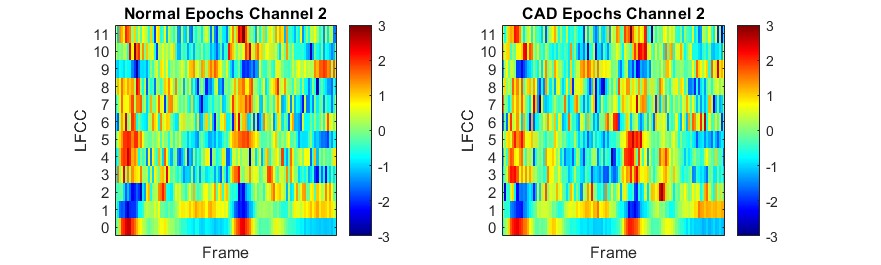}
	    \caption{Average LFCC coefficients of 120 Normal and CAD epochs from channel 2. Each epoch was divided into 108 frames with 50\% overlap}
	    \label{lfcc}
    \end{figure}

\begin{figure}[ht]
        \centering
        \includegraphics[width=16cm]{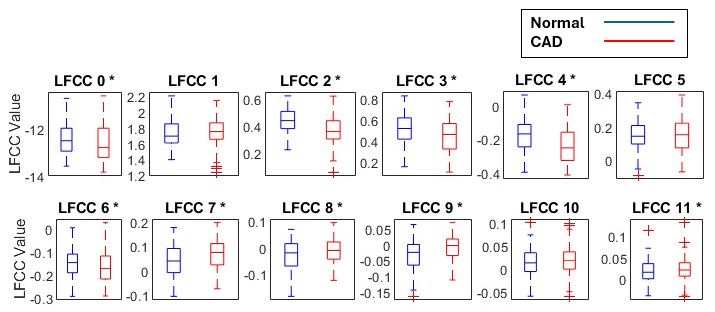}
        \caption{Box plot of LFCCs 0-11 averaged across 108 frames from channel 2. * indicates statistically significant LFCCs, rejecting the null hypothesis that the Normal and CAD samples are from continuous distributions with equal medians.}
        \label{box plot}
    \end{figure}

    Despite the varying lengths of all epochs, each epoch was divided into a predetermined number of frames. As a result, frame sizes were dependent on the subject. This framing technique was implemented to ensure that each frame captured similar cardiac cycle events, regardless of the subject's heart rate. The following frame numbers were investigated with a 50\% overlap: 20-64 and 100-112 in steps of 2. Therefore, for the minimum case of 20 frames with 50\% overlap, the size varied from 100-252.38 ms (with an average of 170.48 ms), and for the maximum case of 112 frames, the variation ranged from 18.92-47.74 ms (with an average of 32.25 ms). A Hanning window was applied to each frame to prevent spectral leakage, as per Equation \ref{hanning}. The power spectrum was computed for each frame by squaring the magnitude spectrum obtained from the Fast Fourier Transform (FFT). A triangular filter bank consisting of 12 linearly spaced filters up to 1 kHz was applied to the power spectrum, yielding a set of filter bank energies. Equation \ref{filter bank} describes the filter bank, where correlation among filter bank energies arises from the overlapping filters.

    \begin{equation} \label{filter bank}
      H_k(f) =
      \begin{cases}
        0, & \text{if } f < f_{k-1} \\
        \frac{f - f_{k-1}}{f_k - f_{k-1}}, & \text{if } f_{k-1} \leq f < f_k \\
        \frac{f_{k+1} - f}{f_{k+1} - f_k}, & \text{if } f_k \leq f < f_{k+1} \\
        0, & \text{if } f \geq f_{k+1}
      \end{cases}
    \end{equation}
    
    \noindent where the frequencies corresponding to the edges of the $k$-th filter are given by $f_{k-1}$, $f_k$, and $f_{k+1}$. The logarithm of the correlated filter bank energies was acquired. To decorrelate the log-filter bank coefficients, denoted by $x_n$, the Discrete Cosine Transformation (DCT) was applied to them, as seen in Equation \ref{DCT}.

    \begin{equation} \label{DCT}
        DCT(i) = \sum_{n=0}^{N-1} x_n \cos \left[ \frac{\pi}{N} i \left(n+\frac{1}{2}\right)\right]
    \end{equation}

    \noindent where $N$ is the number of filter banks (N=12), and $i$ is the DCT coefficient index. A subset of the DCT coefficients was selected to represent the LFCC features utilized in the classification model. Figure \ref{lfcc} displays the average LFCC representation of 120 CAD and 120 Normal epochs from channel 2. Here, coefficients 0-11 are plotted over 108 frames. The LFCC values were averaged across all frames for each epoch to emphasise the statistical significance, and the Wilcoxon rank sum test \cite{wilcoxon1992individual} was applied (p<0.05). Figure \ref{box plot} presents the box plot for each coefficient, with LFCC 0,2,3,4,6,7,8,9 and 11 marked as statistically significant, underscoring the importance of time-varying cepstrum information. The optimal number of coefficients was experimentally determined, where LFCC subsets \{[0-1],[0-2],...,[0-11],[1-2],[1-3],...,[10-11]\} were tested. A feature vector was constructed for each epoch by concatenating the subset of LFCCs from each frame. As a result, the feature dimension was considerably higher than in the previous PSD sub-band investigation. To justify using SVM with RBF kernel, relevant comparisons were made using different classification models, including ANN, k-NN and RF.\\

    ReliefF, a feature selection method that evaluates the importance of features by examining their ability to distinguish between instances of different classes, was utilized to rank the features \cite{robnik2003theoretical}. Predictors are penalised if different values are given between neighbours of the same class, and are rewarded if different values are given between neighbours of opposing classes. 
    Equation \ref{relieff} outlines the method for assigning a weight to feature $j$, with the 100 nearest neighbours, denoted as $\{x_n\}$, utilized in this study.

    \begin{equation}\label{relieff}
        W[j] = \sum_{i=1}^{N} \left( \sum_{\{x_n\} \in C_{x_i}} \frac{-\textrm{diff}(j, x_i, x_n)}{k}.\Tilde{d}_{x_i,x_n} + \sum_{\{x_n\} \notin C_{x_i}} \frac{\textrm{diff}(j, x_i, x_n)}{k}.\Tilde{d}_{x_i,x_n} \right)
    \end{equation}

    \noindent where $W[j]$ is the weight assigned to feature $j$, $N$ is the number of instances, $x_i$ is the $i^{th}$ instance, $C_{x_i} \in [0,1]$ is class of the relevant instance, and $k$ is the number of neighbors. The diff(.) operator computes the absolute difference of feature $j$ between $x_i$ and its neighbour $x_n$, divided by the range of feature values. $\Tilde{d}_{x_i,x_n}$ is the normalised distance between $x_i$ and neighbour $x_n$; thus, further neighbours have less influence on the feature weight. Feature weights can be positive or negative, with the most discriminating feature having the largest positive value. Different feature dimensions comprising the highest-ranked features were examined using an incremental search approach with a step size of 2. For LFCC features, we selected ReliefF because it excels in capturing non-linear relationships between features and the target label. This is crucial in CAD detection, where complex physiological factors can influence PCG signals. ReliefF also outperformed other methods, such as Chi-squared tests or mutual information, which are less robust to noise. Moreover, its computational efficiency is noteworthy, especially in high-dimensional datasets, as it imposes less computational burden than iterative methods like recursive feature elimination.\\
    
    The optimal channel combinations were identified through feature-level fusion, leveraging the performance of the single-channel analysis. Each channel in the combination had its optimal features concatenated into a feature vector. This approach was directly compared with score-level fusion, achieved by averaging the probability estimates of single-channel predictions based on the distance to the hyperplane. We replicated the experiments using alternative filter banks to validate the utilization of a linearly spaced filter bank. Specifically, we compared the LFCC feature set against Gammatone Frequency Cepstral Coefficients (GFCC) with 14 bell-shaped filters, warped to the gammatone scale, and Mel-Frequency Cepstral Coefficients (MFCC) with 12 triangular filters, warped to the mel scale. We assessed both single-channel performance and multi-channel performance through feature-level fusion.

    \subsection{Overview}
\begin{figure}[ht]
        \centering
        \includegraphics[width=\textwidth]{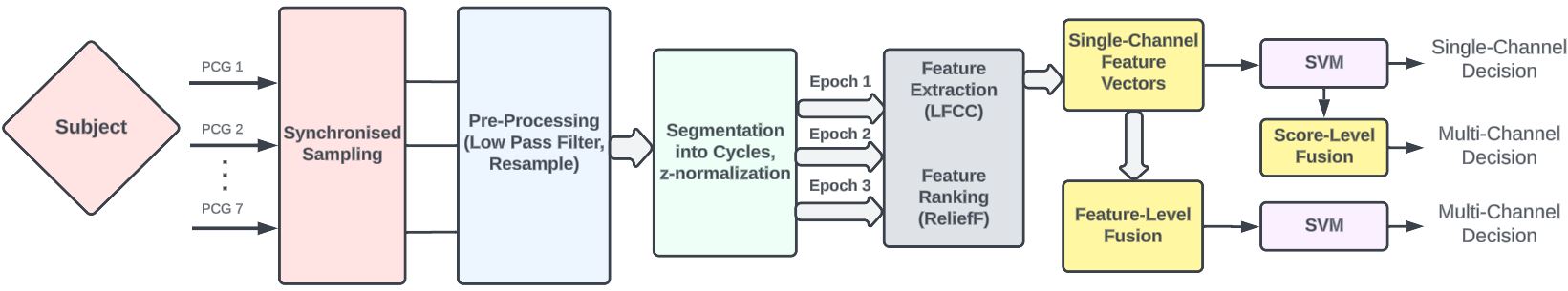}
        \caption{Data processing methodology overview}
        \label{overview}
    \end{figure}

    Figure \ref{overview} illustrates a block diagram summarising the methodology of the LFCC investigation.

    \section{Results and Discussion} \label{results}
    \subsection{Performance of PSD sub-band features}
    Table \ref{psd single channel} presents the SBW, TBW, and FD that yielded the highest-performing metrics for each single channel using SVM with RBF kernel. The table displays both epoch-based metrics and subject-based metrics. All channels produced underwhelming performance individually. From epoch to subject-based metrics, there was an increase in accuracy ranging from 0.56\% to 3.15\%, signifying the importance of integrating multiple samples per subject into the model. Channels 3 and 4 demonstrated the best subject-based performance in accuracy and F1-score. Furthermore, sensitivity surpassed specificity in all channels except 2 and 7. Channel 7 exhibited the poorest performance, registering an F1-score of 48.10\%, the only one to fall below 50\%. Traditionally, heart sounds are not captured from posterior locations because the heart signal can become attenuated due to anatomical factors. Additionally, the back is more prone to muscle activity, which can introduce noise. PSD sub-band features from the back are insufficient for capturing distinct information between the two classes. This limitation can be addressed by extracting features that are more robust to the positional drawbacks. Conversely, Channel 4, with only five features in its classification scheme and a 46.88 Hz SBW, achieved the highest F1-score of 68.35\%. Following closely, Channel 3 attained the second-highest F1-score, utilizing the 59 highest-ranked features from an SBW of 11.72 Hz and a TBW of 700 Hz. These channels are both collected from the subject's left side of the chest, which are traditional auscultation areas. The simplicity of the PSD sub-band features enables channels 3 and 4 to capture discriminating information between CAD and normal cases better than other channels. The most prevalent SBW yielding optimal results was 11.72 Hz, with channels 2, 3, and 5 exhibiting 43, 59, and 55 optimal feature dimensions, respectively. The observed low performance could stem from several factors, including variations in stethoscope placement among subjects and external noise sources in the clinical setting. Additionally, the simplistic nature of the sub-band features might not adequately capture the differences between CAD and non-CAD subjects. It will be shown later that further refinement on feature extraction to include the temporal information such as the LFCC can improve the performance. \\

        \begin{table}[ht]
        \small
        \centering
        \caption{\textbf{Single-Channel Performance using PCG PSD sub-band features}}
        \label{psd single channel}
        \begin{tabular}{llll|llll|llll}
            \hline \hline
            Channel & TBW &SBW  & FD & \multicolumn{4}{c|}{Epoch-based [\%]}  & \multicolumn{4}{c}{Subject-based [\%]}  \\
              &  [Hz] & [Hz] &   & Sens & Spec & Acc  & F1 &Sens & Spec &  Acc & F1   \\
            \hline
            1&500&17.58&25&65.50&58.79&62.15&62.99&68.00&58.75&63.38&64.56 \\
            2&600&11.72&43&63.67&66.17&64.92&64.11&63.50&67.00&65.25&64.11  \\
            3&700&11.72&59&69.00&59.79&64.40&65.63&71.25&59.25&65.25&66.95\\
            \textbf{4}&\textbf{700}&\textbf{46.88}&\b{5}&\b{69.12}&\b{58.46}&\b{63.79}&\b{65.37}&\b{72.38}&\b{61.50}&\b{66.94}&\b{68.35}  \\
            5&800&11.72&55&62.33&54.46&58.40&59.37&64.88&56.88&60.88&61.63  \\
            6&1000&46.88&21&68.17&55.00&61.58&63.65&68.25&55.62&61.94&63.66\\
            7&1000&17.58&56&48.75&55.75&52.25&49.61&46.75&58.88&52.81&48.10\\
            \hline \hline
        \end{tabular}
   \end{table}

   Next, we explored combining features from each channel to assess the robustness of a multi-channel PCG approach. 
   Table \ref{psd multi channel} presents the optimal channel combinations corresponding to each number of channels. Only eight channel combinations achieved above 70\% accuracy and F1-score; 1-3-6, 2-3-6; 1-2-3-6; 1-3-5-6; 1-3-6-7; 1-2-3-5-6; 1-2-3-6-7; and 1-2-3-5-6-7. Channels 3 and 6 are present in all combinations, suggesting discriminative information is captured at the left-fourth and right-second IC. Channel 4, which has 46.88 Hz as its optimal SBW, is not in any combination despite having the best single-channel metrics. No combination had 46.88 Hz as its optimal SBW, offering a possible explanation for why this channel was absent. The combination 1-2-3-6 yielded the highest accuracy and F1-score: 72.25\% and 73.34\%, respectively, which is 5.31\% and 4.99\% greater than the best-performing single-channel case in Table \ref{psd single channel}. This underscores the significance of integrating multi-channel PCG data into the DAQS. Features across different auscultation sites provide richer information that aids classification. 

    \begin{table}[ht]
        \small
        \centering
        \caption{\textbf{Multi-Channel Performance using PCG PSD sub-band features}}
        \label{psd multi channel}
        \begin{tabular}{llll|llll|llll}
            \hline \hline
            Number of & TBW &SBW  & FD & \multicolumn{4}{c|}{Epoch-based [\%]}  & \multicolumn{4}{c}{Subject-based [\%]}  \\
            Channels & [Hz] &[Hz] &   & Sens & Spec & Acc & F1 &Sens & Spec &  Acc & F1  \\
            \hline
            2 (1-6) & 900&11.72&141&68.96&63.38&66.17&66.75&73.75&64.25&69.00&69.85\\
            3 (1-3-6)  &1000&29.3&69&73.21&66.13&69.67&70.64&74.50&69.88&72.19&72.67\\
            \b{4 (1-2-3-6)}&\b{1000}&\b{5.86}&\b{635}&\b{73.62}&\b{63.83}&\b{68.73}&\b{70.16}&\b{77.12}&\b{67.38}&\b{72.25}&\b{73.34}\\
            5 (1-2-3-6-7) &1000&35.16&95&73.08&62.67&67.88&69.33&75.00&66.75&70.88&71.77\\
            6 (1-2-3-5-6-7)&1000&35.16&109&70.08&61.46&65.77&66.69&72.75&65.12&68.94&69.60\\
            7 (1-2-3-4-5-6-7)&300&35.16&45&60.46&63.54&62.00&60.88&62.75&63.62&63.19&62.40\\
            \hline \hline
        \end{tabular}
   \end{table}

    \subsection{Performance of LFCC Features}
    Table \ref{lfcc single channel} presents the performance metrics for single-channel analysis using LFCC features in the 20x5-fold SVM model. It includes the optimal frame number and LFCC subset for each channel. Channel 2 performed the best in this experiment, with 108 frames and LFCCs 0-7 as its optimal configuration. Achieving an accuracy of 73.12\% and an F1-score of 73.33\%, it outperformed other channels by over 4\%. The sensitivity and specificity were 74.62\% and 71.62\%, respectively. The epoch-based accuracy of channel 2 was 68.12\%, again supporting the synopsis of majority voting from multiple subject samples. While Channel 7 continued to exhibit the lowest performance, its accuracy improved significantly by 9.13\% compared to the PSD sub-band feature investigation, aligning it more closely with the performance of other channels. Six out of seven channels showed enhanced performance compared to the simple PSD analysis, signifying the importance of analyzing temporal variations, and PSD shapes comprehensively.\\

    \begin{table}[ht]
        \small
        \centering
        \caption{\textbf{Single-Channel Performance using LFCC features}}
        \label{lfcc single channel}
        \begin{tabular}{llll|llll|llll}
            \hline \hline
            Single & LFCCs & No. & FD & \multicolumn{4}{c|}{Epoch-based [\%]}  & \multicolumn{4}{c}{Subject-based [\%]}  \\
             Channel &  & Frames &   & Sens  & Spec  & Acc   & F1  & Sens  & Spec  &  Acc  & F1   \\
            \hline
            1&0-2&108&317&69.21&64.17&66.69&67.46&71.75&66.25&69.00&69.59 \\
            \b{2}&\b{0-7} &\b{108}&\b{829}&\b{69.63}&\b{66.62}&\b{68.12}&\b{68.37}&\b{74.62}&\b{71.62}&\b{73.12}&\b{73.33}  \\
            3&0-6&28&135&65.08&62.83&63.96&64.02&69.50&65.50&67.50&67.76 \\
            4&0-6&20&69&68.33&63.83&66.08&66.30&72.50&65.38&68.94&69.53 \\
            5&0-7&50&373&60.92&63.12&62.02&60.87&63.38&62.50&62.94&62.12 \\
            6&0-4&20&15&61.25&60.50&60.88&60.64&62.88&60.75&61.81&61.73 \\
            7&0-5&110&597&63.62&57.42&60.52&60.99&65.50&58.38&61.94&62.36 \\
            \hline \hline
        \end{tabular}
   \end{table}

    Combining channels remains essential to achieve clinical significance. Table \ref{lfcc slf} presents the results obtained via score-level fusion. Additionally, Table \ref{lfcc custom channel} showcases the performance metrics achieved through feature-level fusion. The presentation outlines the optimal combination for each channel count, ranging from two channels to the maximum available, seven. 

    \begin{table}[ht]
        \small
        \centering
        \caption{\textbf{Multi-Channel Performance using LFCC Score-Level Fusion}}
        \label{lfcc slf}
        \begin{tabular}{ll|llll|llll}
            \hline \hline
            Number of & FD & \multicolumn{4}{c|}{Epoch-based [\%]}  & \multicolumn{4}{c}{Subject-based [\%]}  \\
            Channels &   &Sens  & Spec  & Acc  & F1  & Sens & Spec  &  Acc  & F1  \\
            \hline
            2 (1-2)&1146&71.21&71.88&71.54&71.41&74.12&73.75&73.94&73.9\\
            3 (2-3-4)&1033&72.88&70.00&71.44&71.47&78.25&71.25&74.75&75.33\\
            4 (1-2-3-7)&1878&73.25&72.21&72.73&72.47&76.62&74.62&75.62&75.4\\
            \b{5 (1-2-3-6-7)}&\b{1893}&\b{73.71}&\b{74.12}&\b{73.92}&\b{73.46}&\b{75.62}&\b{76.88}&\b{76.25}&\b{75.52}\\
            6 (1-2-3-5-6-7)&2266&72.79&72.33&72.56&72.16&74.88&74.75&74.81&74.18\\
            7 (1-2-3-4-5-6-7)&2335&74.46&71.88&73.17&73.06&76.38&74.00&75.19&74.92\\

            \hline \hline
        \end{tabular}
   \end{table}
    \FloatBarrier
    \begin{table}[ht]
        \small
        \centering
        \caption{\textbf{Multi-Channel Performance using LFCC Feature-Level Fusion}}
        \label{lfcc custom channel}
        \begin{tabular}{ll|llll|llll}
            \hline \hline
            Number of & FD & \multicolumn{4}{c|}{Epoch-based [\%]}  & \multicolumn{4}{l}{Subject-based [\%]}  \\
            Channels &   &Sens  & Spec  & Acc  & F1  & Sens & Spec  &  Acc  & F1  \\
            \hline
            2 (2-6)&839&74.08&69.37&71.73&72.23&79.38&73.12&76.25&76.80\\
            
            3 (2-3-7)&1531&76.08&71.29&73.69&73.96&81.00&76.00&78.50&78.69\\
            
            \b{4 (2-3-6-7)}&\b{1681}&\b{79.17}&\b{73.04}&\b{76.10}&\b{76.43}&\b{85.25}&\b{75.62}&\b{80.44}&\b{81.00}\\
            
            5 (1-2-3-6-7)&2093&75.38&73.25&74.31&74.32&80.38&75.88&78.12&78.29\\
            6 (1-2-3-5-6-7)&2281&74.79&73.75&74.27&74.14&79.38&75.62&77.50&77.64\\
            7 (1-2-3-4-5-6-7)&2069&73.83&73.13&73.48&73.30&77.50&75.25&76.38&76.36\\
            
            \hline \hline
        \end{tabular}
   \end{table}

       Feature-level fusion consistently outperforms score-level fusion for all channel counts, demonstrating clinical significance regardless of the total number of channels. The four-channel combination (2-3-6-7) using feature-level fusion achieves an accuracy of 80.44\%, which is 4.82\% higher than the score-level fusion between four channels (1-2-3-7). Additionally, the epoch sensitivity-specificity average is 76.10\%, rendering the DAQS clinically significant both at an epoch and subject level. A five-channel combination (1-2-3-6-7) yielded the highest performance in the score-level fusion method, achieving 76.25\% accuracy. Notably, it was the only combination where the specificity surpassed the sensitivity. In all other scenarios, the sensitivity outweighed specificity. Moreover, all subject-based metrics showed improvement compared to the epoch level. We also note the high FD in each case, resulting from the agglomeration of sensors. This investigation underscores the necessity of multiple sensors to address the challenges a convenient DAQS poses. Compared to the best-performing single channel, there was a 7.32\% increase in accuracy when utilizing four channels via feature-level fusion. Channels 2, 3, 6, and 7 were in the top three performing channel combinations. This study marks the first instance where a PCG acquired from the subject's back (channel 7) has demonstrated effectiveness in CAD classification. The LFCC encapsulates the spectrum shape while incorporating temporal changes through framing, an element not utilized in the previous PSD investigation. In Figure \ref{psd}, we previously observed the average PSD of CAD and non-CAD subjects, clearly illustrating differences in certain frequency bands and distinct shapes. However, the process of LFCC extraction is more complex as it requires additional steps, such as filter bank application and cepstral computation, and typically produces higher-dimensional data because they capture more information about the frequency content. The impact on classification performance is greater, as LFCCs provide a richer signal representation. This helps handle inter-subject variability better, contributing to a more robust and accurate performance. Further, LFCC feature extraction coupled with SVM is not computationally expensive, rendering the system suitable for portable applications. Per previously stated guidelines, the wearable vest demonstrates suitability for practical implementation. Optimal performance is achieved by utilizing only four of the available seven stethoscopes, reducing computational complexity by processing fewer signals and enhancing device portability. Pre-processing and LFCC feature extraction took 0.139 seconds per subject on the computer mentioned in Section \ref{hardware}. We tested the methodology used in \cite{pathak2020detection}, where entropy features are extracted from the SST matrix for each subject across 3 epochs. It took 37 seconds per subject on the same machine—approximately 266 times slower than our method. This highlights the computational efficiency of our proposed method, making the system suitable for mass screening where fast processing is necessary. \\

   The best-performing channel combination underwent testing with various other classifiers. Table \ref{classifier compare} compares the performance among SVM with different kernels, k-NN with varying distance metrics, ANN, and RF. Grid search was conducted on a validation set (24 epochs) to determine the optimal hyperparameters for the k-NN, ANN, and RF classifiers, as detailed in the footnote of Table \ref{classifier compare}. The ANN structure was also tuned using the validation set. Among the tested classification schemes, ANN outperformed all others, achieving a relatively balanced sensitivity and specificity, with accuracy reaching 78.06\%. However, the SVM with the RBF kernel demonstrated the highest performance, as previously reported.

      \begin{table}[ht]
        \small
        \centering
        \caption{\textbf{Performance of the channel combination 2-3-6-7 (feature-level fusion) under various classification schemes.}}
        \label{classifier compare}
        \begin{tabular}{ll|llll|llll}
            \hline \hline
            Classifier & FD &\multicolumn{4}{c|}{Epoch-based [\%]}  & \multicolumn{4}{c}{Subject-based [\%]}  \\
              &&Sens  & Spec  & Acc  & F1  & Sens & Spec  &  Acc  & F1  \\
              \hline
            SVM-LINEAR&1589&71.66&72.00&71.83&71.30&78.00&75.38&76.69&76.22\\
            SVM-CUBIC&775&67.5&78.5&73.00&70.92&70.88&79.50&75.19&73.55\\
            SVM-SIGMOID&1353&71.80&73.17&72.48&71.82&76.88&76.50&76.69&76.08\\
            k-NN EUCLIDEAN& 1421&58.21&71.08&64.64&61.62&59.75&72.50&66.13&62.89 \\
            k-NN COSINE &1771&81.63&60.67&71.15&73.63&84.63&62.50&73.56&76.00 \\
            k-NN CITYBLOCK&1431&57.58&70.58&64.08&61.02&59.63&72.38&66.00&62.95 \\
            k-NN CORRELATION&1801&80.88&59.71&70.29&73.01&85.75&61.25&73.50&76.20 \\
            ANN&1501&71.54&75.08&73.31&72.32&77.75&78.38&78.06&77.25\\
            RF&1101&71.25&71.38&71.31&70.68&74.12&74.00&74.06&73.24 \\
            \b{SVM-RBF (Proposed)}&\b{1681}&\b{79.17}&\b{73.04}&\b{76.10}&\b{76.43}&\b{85.25}&\b{75.62}&\b{80.44}&\b{81.00}\\ 
            \hline \hline
        \end{tabular}
        \begin{tablenotes} \scriptsize
             \item[*] \textit{k-NN}: 11 neighbours. \textit{ANN}: No. hidden layers: 1; hidden nodes: 700; Activation Function: ReLU, Sigmoid; learning-rate: 0.0001; Adam Optimizer; Batch Size: 64; Epochs: 150; L2 Regularization: 0.01; Loss Function: Binary Cross Entropy. \textit{RF}: Criterion: Gini Impurity; No. Estimators: 500; Max depth: 100; Max features: `sqrt': Min samples/split: 6; Max leaf nodes: 20
        \end{tablenotes}
   \end{table}

        \newpage
    \section{Comparison with GFCC and MFCC} \label{compare} 

   We systematically reproduced previous experiments using GFCC and MFCC-based features, conducting a comparative analysis to justify our selection of LFCC as the primary feature set. Tables \ref{gfcc single channel} and \ref{mfcc single channel} present the single-channel performance results using GFCC and MFCC features, respectively, indicating the optimal cepstral coefficients and frame numbers. Tables \ref{gfcc custom channel} and \ref{mfcc custom channel} display the multi-channel performance of GFCC and MFCC feature sets, respectively, showcasing the best channel combinations for each total number of channels, achieved through feature-level fusion of each singular channel's optimal counterparts. \\

   \begin{table}[ht]
        \small
        \centering
        \caption{\textbf{Single-Channel Performance using GFCC features}}
        \label{gfcc single channel}
        \begin{tabular}{llll|llll|llll}
            \hline \hline
            Single & GFCCs & No. & FD & \multicolumn{4}{c|}{Epoch-based [\%]}  & \multicolumn{4}{c}{Subject-based [\%]}  \\
             Channel &  & Frames &   & Sens  & Spec  & Acc  & F1  & Sens  & Spec  &  Acc  & F1   \\
            \hline
            1&2-10&100&871&67.42&65.67&66.54&66.55&70.62&69.38&70.00&69.61\\
            2&0-11&100&1187&69.67&64.00&66.83&67.30&73.38&65.00&69.19&70.01\\
            3&1-6&24&143&63.12&65.17&64.15&63.57&65.62&68.38&67.00&66.19\\
            4&0-6&30&59&68.92&66.00&67.46&67.34&70.00&68.25&69.12&68.81\\
            5&0-6&64&439&65.79&62.42&64.10&64.01&68.62&63.00&65.81&65.88\\
            \b{6}&\b{2-7}&\b{20}&\b{109}&\b{70.33}&\b{67.79}&\b{69.06}&\b{69.23}&\b{72.50}&\b{68.38}&\b{70.44}&\b{70.77}\\
            7&0-8&108&953&63.33&62.83&63.08&62.36&66.38&65.25&65.81&64.96\\
            \hline \hline
        \end{tabular}
   \end{table}
   \vspace{-3mm}
     \begin{table}[ht]
        \small
        \centering
        \caption{\textbf{Single-Channel Performance using MFCC features}}
        \label{mfcc single channel}
        \begin{tabular}{llll|llll|llll}
            \hline \hline
            Single & MFCCs & No. & FD & \multicolumn{4}{c|}{Epoch-based [\%]}  & \multicolumn{4}{c}{Subject-based [\%]}  \\
             Channel &  & Frames &   & Sens  & Spec  & Acc  & F1  & Sens  & Spec  &  Acc  & F1   \\
            \hline
            \b{1}&\b{0-2}&\b{24}&\b{55}&\b{70.62}&\b{61.00}&\b{65.81}&\b{67.35}&\b{75.75}&\b{64.50}&\b{70.12}&\b{71.64}\\
            2&0-6&108&756&68.33&66.46&67.40&67.16&70.25&68.62&69.44&69.12\\
            3&0-4&28&135&66.58&62.88&64.73&65.13&69.75&67.38&68.56&68.70\\
            4&0-3&28&111&67.75&67.79&67.77&67.05&68.12&70.25&69.19&68.20\\
            5&0-9&64&631&63.58&61.88&62.73&62.33&64.75&62.50&63.62&63.28\\
            6&0-2&62&123&62.83&62.46&62.65&62.09&64.88&64.88&64.88&64.11\\
            7&0-10&62&675&63.54&57.21&60.38&61.06&68.25&58.62&63.44&64.61\\
            \hline \hline
        \end{tabular}
   \end{table}
   In our assessment of individual channel performance, we consistently observed that both the GFCC and MFCC feature sets outperformed the LFCC feature set across the majority of channels. Notably, exceptions were found in channel 2 for both GFCC and MFCC, and in channel 3 specifically for the GFCC scenario. Significant improvements were particularly notable in channels 6 and 7. Channel 6 demonstrated exceptional performance, with GFCC features yielding a subject-level accuracy of 70.44\%, significantly surpassing the 61.81\% achieved using LFCC features, thus establishing it as the highest-performing channel. Similarly, in channel 7, GFCC features achieved the highest accuracy of 65.81\% among all experiments conducted. For the MFCC feature set, channel 1 exhibited the most promising results, achieving an accuracy of 70.12\%, an improvement of 1.12\% over the LFCC case. However, both GFCC and MFCC features fell short in channel 2, with subject-level accuracies lower by 3.93\% and 3.68\%, respectively, compared to LFCC. 
   
   \FloatBarrier
   \begin{table}[ht]
        \small
        \centering
        \caption{\textbf{Multi-Channel Performance using GFCC Feature-Level Fusion}}
        \label{gfcc custom channel}
        \begin{tabular}{ll|llll|llll}
            \hline \hline
            Number of & FD & \multicolumn{4}{c|}{Epoch-based [\%]}  & \multicolumn{4}{c}{Subject-based [\%]}  \\
            Channels &   &Sens  & Spec  & Acc  & F1  & Sens & Spec  &  Acc  & F1  \\
            \hline
            2 (1-7)&1871&72.25&67.42&69.83&70.02&75.88&71.00&73.44&73.30\\
            3 (1-3-7)&2016&74.79&68.42&71.60&71.95&77.62&71.50&74.56&74.62\\
            \b{4 (1-3-6-7)}&\b{2121}&\b{74.75}&\b{68.12}&\b{71.44}&\b{71.83}&\b{78.75}&\b{70.88}&\b{74.81}&\b{75.09}\\
            5 (1-3-5-6-7)&2266&73.29&68.96&71.13&71.21&77.00&70.62&73.81&74.02\\
            6 (1-2-3-5-6-7)&3416&73.37&69.62&71.50&71.57&75.00&71.75&73.38&73.21\\
            7 (1-2-3-4-5-6-7)&3961&72.29&68.50&70.40&70.42&74.12&69.75&71.94&71.94\\
            
            \hline \hline
        \end{tabular}
   \end{table}

   \begin{table}[!h]
        \small
        \centering
        \caption{\textbf{Multi-Channel Performance using MFCC Feature-Level Fusion}}
        \label{mfcc custom channel}
        \begin{tabular}{ll|llll|llll}
            \hline \hline
            Number of & FD & \multicolumn{4}{c|}{Epoch-based [\%]}  & \multicolumn{4}{c}{Subject-based [\%]}  \\
            Channels &   &Sens  & Spec  & Acc  & F1  & Sens & Spec  &  Acc  & F1  \\
            \hline
            2 (1-2)&827&72.54&69.58&71.06&71.20&73.62&71.38&72.50&72.41\\
            3 (1-2-5)&1251&72.12&70.62&71.38&71.30&73.50&73.88&73.69&73.27\\
            4 (2-3-6-7)&1566&77.38&72.50&74.94&75.10&79.75&72.50&76.12&76.45\\
            \b{5 (1-2-3-6-7)}&\b{1774}&\b{77.54}&\b{72.92}&\b{75.23}&\b{75.53}&\b{79.25}&\b{74.75}&\b{77.00}&\b{77.09}\\
            6 (1-2-3-5-6-7)&2216&76.42&73.67&75.04&74.95&77.75&75.12&76.44&76.19\\
            7 (1-2-3-4-5-6-7)&2441&74.67&72.00&73.33&73.14&75.50&73.75&74.62&74.09\\
            
            \hline \hline
        \end{tabular}
   \end{table}

   Despite the promising performance observed in single-channel analysis, the results from multi-channel metrics obtained through feature-level fusion were underwhelming when compared to LFCC. None of the combinations yielded clinically significant results with GFCC features. Combining channels 1-3-6-7 produced the highest subject-level accuracy and F1-score of 74.81\% and 75.09\%, respectively. However, the utilization of MFCC features showed improved results, with the combination 1-2-3-6-7 demonstrating the highest performance. This combination achieved epoch-level and subject-level accuracies of 75.23\% and 77\%, respectively, indicating clinical significance at both levels. Nevertheless, these achievements did not match the performance attained with the LFCC feature set. \\

    Channels 3, 6, and 7 consistently appear in all the top-performing channel combinations across LFCC, GFCC, and MFCC feature sets, indicating their significance. Channel 1 is featured in the GFCC and MFCC top combination, while channel 2 is found in the LFCC and MFCC top combination, suggesting their importance. Conversely, channels 4 and 5 are less important based on our findings. The approximate positions of channels 3 (left fourth IC space) and 6 (right second IC space), which align with the heart's natural slant, could lead to richer signals stemming from these sites. Although the heart is predominantly on the left side of the body, the right atrium slightly overlaps the right side, where channel 6 captures heart sounds. Additionally, the left side of the heart is positioned more posteriorly, where the left circumflex coronary artery wraps around, potentially explaining the significance of channel 7 (posterior position) in classification. Channels 1 (left midaxillary line) and 2 (near the apex) are also directly over the heart's natural position in a sitting posture. Channels 4 (left second IC space) and 5 (right fourth IC space) are not positioned directly over the heart, offering an explanation as to why these channels are less important. This aligns with the findings in \cite{pathak2020detection}, where the lowest single-channel performance was observed in the left second IC space out of four channels on the left side of the chest.

   \section{Observations on subjects with PTCA or CABG} \label{Additional}
    Our data set includes additional subjects that were excluded from the initial study, comprising six individuals who have undergone PTCA and four who have had CABG. \cite{dragomir2016acoustic} computed power ratios for CAD, Normal, and post-PTCA subjects by determining the ratio of total power above to below 150 Hz from the FFT of the PCG. Their analysis revealed no statistically significant difference between patients post-stenting and those without coronary stenosis, indicating a reduction in murmur sounds. In CABG, occluded arteries may still exist; however, blood flow to the myocardium is redirected through the new arteries, reducing flow through the occluded areas \cite{mullany2003coronary}. Consequently, we anticipate a decrease in murmurs. Ideally, our model would classify post-CABG and PTCA subjects as Normal, assuming they have no restenosis. However, training a model that includes these subject types is essential in practice. Nevertheless, we have only assessed our best-performing model (LFCC PCG, channel combination 2,3,6,7) on these subjects to demonstrate the potential for post-monitoring with the wearable vest. We trained the model using the entire dataset, comprising 120 CAD and 120 Normal epochs. Subsequently, we applied the trained SVM model to post-procedural subjects, with preprocessing identical to that described in Section \ref{preprocessing section}. Table \ref{Post operative} displays the results and compares them to the ideal outcome. \\
    
    \begin{table}[ht]
        \small
        \centering
        \caption{\textbf{System evaluation on subjects previously undergone PTCA or CABG}}
        \label{Post operative}
        \begin{tabular}{c|cccc}
            \hline \hline
            Subject & \makecell{Previous Procedure} & New Diagnosis & Ideal Outcome & \makecell{Model Prediction} \\
            \hline
            1 & CABG & No restenosis & Normal  & Normal\\
            2 & 3x CABG & No restenosis & Normal & Normal\\
            3 & CABG & Post CABG Stenosis & CAD & CAD\\
            4 & CABG & Post CABG Stenosis & CAD & Normal \\
            5 & LAD PTCA & No restenosis & Normal & CAD\\
            6 & LAD PTCA & In-Stent Restenosis & CAD & CAD\\
            7 & 3x PTCA & TVCAD & CAD & CAD \\
            8 & PTCA & SVCAD & CAD & CAD  \\
            9 & PTCA & DVCAD & CAD & Normal  \\
            10 & 2x PTCA & TVCAD & CAD & CAD\\

            \hline \hline
 
        \end{tabular}
   \end{table}

    Out of ten additional subjects, our best-performing model predicted seven cases that matched the ideal outcome. While this test set was relatively small, it highlights the potential of the wearable vest in post-procedural monitoring, particularly in flagging individuals where restenosis has occurred after CABG or PTCA. However, there were two cases in which the model failed to predict restenosis: one CABG and one PTCA subject. This may be attributed to the fact that our model achieved slightly above 80\% accuracy and that post-procedural subjects were not included in the model training. Moving forward, we aim to collect data from a larger sample of such subjects to assess post-procedural monitoring capabilities better.

    \section{Comparison with Existing Literature} \label{compare to lit}
    As we have collected a new and realistic dataset, comparing performance metrics with other studies that prioritize classification accuracy using different databases is not fair. Nevertheless, Table \ref{compareRes} compares our subject-based metrics with those reported in the literature. We emphasize that the goal of this study is not to surpass the performance of existing studies but rather to analyze data recorded under realistic clinical settings in an easy and convenient manner. We outline the disadvantages of each existing study and assess their practicality for real-world implementation.

    \begin{table}
    \small
        \centering
        \vspace{-3mm}
        \caption{\textbf{Performance comparison with existing studies}}
        \label{compareRes}
        \begin{tabular}{c|ccc p{5cm}}
            \hline \hline
            Study & Database & Feature Type,  Classifier &Result [\%] & Remarks\\
            \hline
            Makaryus \textit{et al.} \cite{makaryus2013utility}&\makecell{19 CAD\\ 142 normal}  & \makecell{Microbruit score, \\ Logistic Regression} &\makecell{Sens: 89.50 \\ Spec: 57.70 \\ Acc: 64.50} &  \makecell[l]{ \fs- Results are not clinically significant \\ \fs Concl: Not practical \ding{55}} \\
           & \\
            Schmidt \textit{et al.} \cite{schmidt2015acoustic}& \makecell{63 CAD\\ 70 normal}  & \makecell{Frequency and \\ nonlinear features, \\ multivariate classifier}&\makecell{Sens: 72.00 \\ Spec: 65.20 \\ Acc: 68.40} & \makecell[l]{\fs - Results are not clinically significant \\\fs Concl: Not practical \ding{55}}\\
            &\\
            Li \textit{et al.} \cite{li2020fusion}$^1$ & \makecell{120 CAD\\ 55 normal} & \makecell{Multi-domain \\ features and deep \\ learning features, \\ MLP} &\makecell{Sens: 93.00 \\ Spec: 83.40 \\ Acc: 90.40}  & \makecell[l]{\fs- Unrealistic patient preparation \\ \fs\; (15 min in temp controlled room) \\ \fs
            - Precise sensor position, supine  \\\fs- Quiet, non-clinical setting \\\fs- High cost \\ \fs Concl: Not practical \ding{55}} \\
            &\\
            Iqtidar \textit{et al.} \cite{iqtidar2021phonocardiogram}$^1$& \makecell{78 CAD \\ 75 normal}  & \makecell{MFCC and 1D-ALTP \\ features, SVM-cubic}&\makecell{Sens: 98.20 \\ Spec: 93.50 \\ Acc: 98.30} & \makecell[l]{\fs - Precise sensor placement \\\fs- Only CAD group collected \\\fs \; in clinical setting \\\fs- Angiogram not used for labelling  (PCG \\ \fs \; signals labelled) \\\fs- Samples from same subject can appear \\ \fs \; in train and test groups \\\fs Concl: Not practical \ding{55}}\\
            &\\
            Huang \textit{et al.} \cite{huang2022customized}$^1$ & \makecell{206 CAD \\ 348 normal}  & \makecell{MFCC deep learning \\ (CNN+LSTM), RF }&\makecell{Sens: 96.12 \\ Spec: 96.12 \\ Acc: 96.05} & \makecell[l]{\fs- Unrealistic patient preparation \\ \fs\; (15 min in temp controlled room) \\ \fs
            - Precise sensor position, supine  \\\fs- Quiet, non-clinical setting \\\fs- High cost \\ \fs Concl: Not practical \ding{55}}\\
            
            &\\
            Pathak \textit{et al.} \cite{pathak2020detection}& \makecell{40 CAD \\ 40 normal}  & \makecell{Entropy features \\ from SST matrix, \\ SVM-linear}&\makecell{Sens: 84.38 \\ Spec: 85.25 \\ Acc: 84.81} & \makecell[l]{ \fs - Four stethoscopes held with tape \\ \fs- Precise sensor position, supine  \\\fs- Quiet, non-clinical setting  \\ \fs Concl: Not practical \ding{55}}\\ 
            &\\
            Liu \textit{et al.} \cite{liu2021detection}$^1$ & \makecell{21 CAD \\15 normal} & \
            \makecell{Multi-domain \\ features with entropy \\ and XEntropy, \\ SVM-RBF}&\makecell{Sens: 88.00 \\ Spec: 93.00 \\ Acc: 90.92}& \makecell[l]{\fs - Five precise sensor positions, supine  \\\fs- Non-clinical setting \\ \fs - All CAD subjects have left anterior \\ \fs \; descending stenosis (system may not \\ \fs \;generalise well to right artery CAD) \\ \fs Concl: Not practical \ding{55}} \\
            &\\
            This Study$^2$& \makecell{40 CAD \\ 40 normal} & \makecell{LFCC, SVM-RBF}&\makecell{Sens: 85.25 \\ Spec: 75.62 \\ Acc: 80.44} & \makecell[l]{ \fs - No patient preparation \\ \fs - No specialist training required \\ \fs - No precise stethoscope placement \\ \fs - Easy and convenient wearable vest \\ \fs- Seated (no hospital bed required) \\ \fs - Clinical environment \\ \fs - Low Cost \\ \fs - Post procedural monitoring demonstrated\\ \fs  \b{Concl: Suitable for real-life use} \checkmark}\\
            \hline
            
        \end{tabular}
        \begin{tablenotes} \scriptsize
             \item[*] $^1$Subject-level accuracy not reported. Fragment/epoch-level metrics reported.
             \item[**] $^2$ Our method can be implemented in a real-life and non-controlled environment. Hence, we expect our results to underperform those of existing studies that were conducted in a controlled and non-realistic manner.
        \end{tablenotes}
        
   \end{table}
   
    \section{Practical Implementation} \label{practical}
    Implementing the wearable vest in a real-world scenario will incorporate challenges. We outline the main challenges and outline how the system can adapt to provide solutions in Table \ref{challenge}.

    \begin{table}[h]
    \small
        \centering
        \caption{\b{Challenges/limitations and solutions for real-world implementation of wearable PCG vest}}
        \begin{tabular}{p{4cm}|p{5cm}|p{5cm}}
        \hline
        \hline
             \b{Challenge/limitation} & \b{Description} & \b{How the system can adapt} \\
             \hline
             Variations in stethoscope placement& Ensuring consistent and accurate placement of multiple stethoscopes across different subjects and sessions can be challenging. Variations in placement can lead to inconsistencies in signal quality and affect the model's performance. Although LFCC features produce clinically significant results, extreme position differences can lead to poor generalisation & \begin{itemize}[leftmargin=*]
                 \item Ergonomic designs to ensure standard placements can be incorporated. \item Feedback mechanisms or indicators to assist users in achieving optimal sensor placement can be implemented.
                 \item Machine learning models can be trained that are robust to sensor placement.
             \end{itemize} \\
             \hline
             Interference and external noise& PCG signals can be corrupted by external noise sources, including ambient room noise, movement artifacts, and physiological factors. Poor signal-to-noise ratio can result in erroneous feature extraction and affect the reliability of the classification model. & \begin{itemize}[leftmargin=*]
                 \item Implement real-time monitoring and adjustment to enhance signal quality during acquisition.
             \end{itemize}\\
             \hline
             User comfort and compliance & Wearable devices must be comfortable. Vest fit, material comfort, and sensor size are factors that can influence user compliance and data quality. & \begin{itemize}[leftmargin=*]
                \item Design wearable with a Human Factors and Ergonomics (HFE) team that focuses on designing products that optimize user comfort, efficiency and safety
                 \item Incorporate user-feedback schemes to continuously improve user experience
             \end{itemize}\\
             \hline
             \hline 
        \end{tabular}
        
        \label{challenge}
    \end{table}
    \section{Conclusion} \label{Summary}
    This study has pushed the boundaries of state-of-the-art DAQS and data collection methodologies by employing a wearable vest equipped with seven electronic stethoscopes all synchronously sampling from the subject. The process of fitting the vest, testing the connections, recording a 10-second measurement, and removing the vest can be completed in under two minutes, emphasising the ease and convenience of the system. This contrasts existing literature, where laborious subject preparation, precise sensor placement, and stringent environmental conditions make it difficult for practical implementation. While the wearable vest offers remarkable ease of use, it also presents challenges. As the sensors are fixed to the vest, variations in body shapes can lead to inter-subject placement variability. Additionally, the clinical environment in which data collection takes place introduces noise into the system. Despite these obstacles, the system has demonstrated performance suitable for practical use, aiming to achieve a sensitivity-specificity average of greater than 75\%. \\

    The initial investigation of PSD sub-band features proved ineffective in meeting the target threshold. Various sub-band widths across specified total bandwidths were extracted and utilized as features in a 20x5-fold cross-validation SVM with RBF kernel. Employing MRMR, the features were ranked to identify the most relevant ones. The optimal channel combination was 1-2-3-6, achieving an accuracy of 72.25\% and an F1-score of 73.34\%. To enhance the results, features with higher complexity were required. LFCC features were extracted, where each single channel's optimal frame number and LFCC subset were established through the same 20x5-fold SVM model. These optimum features from each channel were concatenated to study the performance of combining channels at a feature level. A combination of 2-3-6-7 produced the best performance metrics and verified the system's practical viability. It was the first study to show the effectiveness of PCG signals acquired from the back in CAD detection. The sensitivity-specificity average was 80.44\%, and the F1-score was 81.00\%. The epoch-based accuracy and F1-score were 76.10\% and 76.43\%, respectively, accentuating the need for multiple samples per subject. Moreover, it was found that channels 3, 6, and 7 consistently appear in the most effective combinations for LFCC, GFCC, and MFCC feature sets, underscoring the importance of these auscultation areas. The best-performing, clinically significant model was evaluated on post-PTCA and post-CABG subjects, with initial results suggesting potential for post-procedural monitoring.

    \section{Future Direction} \label{Future}
    The popularity of deep and transfer learning involving biosignals has recently grown. This study focused on hand-crafted features due to the small data set available and the desire to offer greater interpretability and explainability to the obtained results. As more data is collected using the wearable vest, there will be ample opportunity to delve into deep and transfer learning techniques, leveraging the expanded dataset for more effective exploration and analysis. A synchronous electrocardiogram signal that can be captured with the multi-channel PCG data will be investigated in future work. This can assist heart cycle segmentation and provide more discriminating information to aid classification. \\

    Currently, all CAD-positive subjects who participated in the study presented with symptoms. In the future, we aim to conduct a study on detecting CAD in asymptomatic patients, which is the ultimate goal of the pre-screening device. Additionally, we will extend the study to incorporate female subjects.

    \section*{Authors' contribution}
    Matthew Fynn: conceptualization, methodology, software, validation, formal analysis, investigation-data collection, writing-original draft, writing-review \& editing, visualization. Kayapanda Mandana: investigation-data collection, data curation. Javed Rashid: investigation-data collection, writing-review \& editing. Sven Nordholm: resources, project administration, writing-review \& editing, supervision. Yue Rong: resources, writing-review \& editing, supervision. Goutam Saha: resources, project administration, writing-review \& editing, supervision.

    \section*{Ethics approval and consent}
    The study received approval from the ethics committee of Fortis Hospital, Kolkata, India, where the data collection took place. Informed consent was obtained from all participating subjects.
    
    \section*{Acknowledgement}
    We thank \textit{Ticking Hearty Pty Ltd} for providing their wearable vest design for data collection. We thank Milan Marocchi, Arnab Maity, and Sudestna Nahak for their valuable remarks and feedback towards this study.
    
    \section*{Conflict of Interest}
    We declare that we have no conflicts of interest.

 \bibliographystyle{elsarticle-num} 
 \bibliography{cas-refs}





\end{document}